\documentclass[
  reprint,
  aps,
  prd,
  nofootinbib,
  superscriptaddress
]{revtex4-1}

\usepackage{amssymb,amsmath,amstext,amsfonts}

\usepackage{bm}
\usepackage{graphicx}
\usepackage{xcolor}
\usepackage{lipsum}
\usepackage{braket}
\usepackage{comment}
\usepackage{hyperref}
\hypersetup{hidelinks,
  pdftitle={Hubble Susceptibility: Beyond Effective-Mass Spectroscopy in the Cosmological Collider},
  pdfauthor={Shuntaro Aoki and Masahide Yamaguchi}}
\usepackage{mathrsfs}

\newif\ifshowchanges

\begin{document}

\title{Hubble Susceptibility:\texorpdfstring{\\}{ }
Beyond Effective-Mass Spectroscopy in the Cosmological Collider}

\author{Shuntaro Aoki}
\affiliation{
  RIKEN Center for Interdisciplinary Theoretical and Mathematical Sciences
  (iTHEMS),\protect\\
  Wako, Saitama 351-0198, Japan
}

\author{Masahide Yamaguchi}
\affiliation{
   Cosmology, Gravity and Astroparticle Physics Group,
  Center for Theoretical Physics of the Universe,\protect\\
  Institute for Basic Science, 
  Daejeon 34126, Korea
}
\affiliation{
  \mbox{Department of Physics, Institute of Science Tokyo,
  2-12-1 Ookayama, Meguro-ku, Tokyo 152-8551, Japan
}}
\affiliation{   \mbox{ Department of Physics and IPAP, Yonsei University,
  50 Yonsei-ro, Seodaemun-gu, Seoul 03722, Korea}
}

\begin{abstract}
In this note, we introduce the Hubble susceptibility of
cosmological-collider masses, encoded in the scale dependence of the
squeezed-limit oscillation. Heavy fields generically receive
Hubble-induced mass corrections during inflation, and we show that the
local oscillation frequency and its drift can disentangle such corrections
from an intrinsic mass contribution at leading slow-roll and adiabatic order.
\end{abstract}

\maketitle

\section{Introduction}

The cosmological collider provides a unique opportunity to probe particles
with masses comparable to the Hubble scale during inflation~\cite{Chen:2009zp,Baumann:2011nk,Noumi:2012vr,Arkani-Hamed:2015bza}.  In particular,
nonanalytic momentum dependence and characteristic oscillations in primordial
correlators can reveal information about the masses and spins of particles
present in the inflationary era.  Such a detection would already constitute
valuable evidence for physics beyond the Standard Model.

The mass inferred from a cosmological-collider signal, however, is the
effective mass during inflation. A simple and well-motivated form of the inflationary effective mass is
\begin{equation}
  m_{\rm eff}^2
  =
  m_0^2+cH^2 ,
  \label{m_eff0}
\end{equation}
where \(m_0\) denotes an intrinsic (Hubble-independent) mass and \(cH^2\)
represents the contribution induced by the inflationary background.
Hubble-induced masses arise in a variety of settings, including non-minimal
couplings to gravity~\cite{Kumar:2017ecc}, supergravity
constructions~\cite{Stewart:1994ts,Yamaguchi:2011kg}, and loop-induced
corrections~\cite{Chen:2016nrs}.  Throughout the main text, \(m_0^2\) and \(c\)
are assumed to be constant along the inflationary background.  The
effective mass measured during inflation therefore need not coincide with
the intrinsic mass parameter \(m_0\), which in turn need not equal the
present-day pole mass.

From a particle-physics perspective, it is desirable not only to establish
the presence of a new particle, but also to separate its intrinsic mass scale
from the contribution generated by the inflationary environment.  This
separation is impossible from a measurement of \(m_{\rm eff}\) at a single
Hubble scale alone.  One may attempt to use mass splittings between multiple
particles, for which a universal Hubble correction can cancel.  As we discuss
below, however, this possibility relies on additional assumptions and the
corresponding mass-difference signal is not guaranteed to be observable.

Scale-dependent collider signals from time-dependent masses have been
studied in Refs.~\cite{Wang:2019curved,
Reece:2022largefield,Aoki:2023dynamical}. Building on these studies, we
focus on the response to the slow evolution of the Hubble scale during
inflation. Since the relative importance of \(m_0^2\) and \(cH^2\)
changes as \(H\) evolves, the cosmological-collider frequency acquires a small
scale dependence.  This response is encoded in the accumulated phase of the
squeezed-limit oscillation and appears as a chirp of its otherwise
log-periodic waveform.  We refer to the response of the collider frequency
to the Hubble scale as the \emph{Hubble susceptibility}.  Within the mass
ansatz~\eqref{m_eff0}, the local oscillation frequency and its drift allow the
Hubble-independent and Hubble-induced contributions to be disentangled.

We first discuss the limitations of mass-splitting spectroscopy in Sec.~\ref{sec:mass-splitting} and then
derive the relation between the collider phase and the Hubble susceptibility in Sec.~\ref{sec:hubble-susceptibility}.  The appendices contain the exact scalar-loop result relevant to the
mass-splitting discussion, the extension beyond the minimal mass ansatz,
and the WKB derivation together with corrections to the leading phase
relations in Appendices~\ref{app}, \ref{app:general-mass}, and
\ref{app:phase-corrections}, respectively.
%----------------------------------------
%----------------------------------------
\section{Limitations of mass-splitting spectroscopy}
\label{sec:mass-splitting}

We first revisit the mass-difference signal identified in
Ref.~\cite{Aoki:2020zbj}, where the one-loop bispectrum generated by two
heavy scalar fields was studied.  More specifically, we consider the mixed
scalar-bubble diagram in which \(\sigma_1\) and \(\sigma_2\) circulate in
the loop, with the two hard external inflaton modes attached to a quartic
vertex and the soft mode attached to a cubic vertex.  It was shown that,
for nearly degenerate masses, this contribution can contain a long-period
oscillation controlled by the mass difference. This
observation suggests a possible way to probe mass splittings in the
inflationary particle spectrum, which we refer to here as
mass-splitting spectroscopy.

In the squeezed limit, the part of the signal involving both massive species
schematically takes the form
\begin{align}
\nonumber S_{\rm osc}^{(12)}
  \sim &\ 
  C_+ \cos\!\left[(\mu_1+\mu_2)L+\delta_+\right]
  \\
  &+
  C_- \cos\!\left[(\mu_1-\mu_2)L+\delta_-\right],
  \label{eq:loop-mass-splitting}
\end{align}
where \begin{equation}
  \mu_i
  \equiv
  \sqrt{\frac{m_{i,\rm eff}^2}{H^2}-\frac94},\quad m_{i, \rm eff}^2
  \equiv 
  m_{i,0}^2+c_iH^2,\label{mu_i}
\end{equation}
with $i=1,2$. Here \(\mu_i\) is the de Sitter mass parameter of \(\sigma_i\), which is assumed to be real and positive.  We denote
the short- and long-mode momenta by \(k_S\) and \(k_L\), respectively, with
\(k_L\ll k_S\), and define
\begin{equation}
  L\equiv\ln\frac{k_S}{k_L}.\label{def_L}
\end{equation}
Without loss of generality, we take \(\mu_1\geq\mu_2\). When \(\mu_1\simeq\mu_2\), the second term gives a long-wavelength modulation and can be useful for diagnosing a nearly degenerate spectrum.
However, the detection of such a frequency does not by itself provide
a model-independent measurement of the intrinsic masses $m_{i,0}$.

There is, nevertheless, an idealized situation worth noting.  In a
minimal supergravity setup with a canonical K\"ahler potential, scalar
fields $\sigma_i$ can acquire a universal Hubble-induced mass contribution,
\begin{equation}
  c_1=c_2=3 ,
\end{equation}
as discussed in Ref.~\cite{Yamaguchi:2011kg}.  In this case, the collider
frequencies~\eqref{mu_i} become
\begin{equation}
  \mu_i
  =\sqrt{\frac34+\frac{m_{i,0}^2}{H^2}} .
\end{equation}
When the intrinsic-mass contributions are subleading,
\(m_{i,0}^2/H^2\ll 3/4\), the difference frequency becomes
\begin{equation}
  \mu_1-\mu_2
  \simeq
\frac{m_{1,0}^2-m_{2,0}^2}{\sqrt{3}H^2}.
  \label{eq:bare-mass-splitting}
\end{equation}
Since the common \(3H^2\) contribution cancels from the effective
squared-mass difference, the observed frequency splitting can be
translated into the intrinsic squared-mass splitting \(m_{1,0}^2-m_{2,0}^2\) at
leading order.
More generally, for any common \(c_1=c_2\), the exact identity is
\((\mu_1+\mu_2)(\mu_1-\mu_2)=(m_{1,0}^2-m_{2,0}^2)/H^2\).

This inference, however, relies crucially on the universality of the
Hubble-induced contribution.  Higher order K\"ahler corrections can
easily spoil this universality~\cite{Yamaguchi:2011kg}, in which case the
difference frequency is sensitive not only to \(m_{1,0}^2-m_{2,0}^2\), but also
to \(c_1-c_2\).

Moreover, the visibility of the difference-frequency component depends
on the interactions.  The late-time analysis of Ref.~\cite{Aoki:2020zbj}
found sum-frequency dominance for the derivative interaction considered
there and the opposite hierarchy for the non-derivative interaction near mass degeneracy.
Recent progress in the exact analytic evaluation of loop
correlators~\cite{Xianyu:2022jwk,Qin:2024gtr} allows us to revisit these
conclusions without relying on the late-time approximation.  We find that
the qualitative hierarchy for the representative time-derivative
contribution remains unchanged near mass degeneracy.  By contrast, the conclusion for the
non-derivative interaction must be revised: the nonanalytic oscillatory
signal vanishes, as shown in Appendix~\ref{app}.  This provides the scalar
counterpart of the fermion-bubble Yukawa null found in
Ref.~\cite{Aoki:2026vbc}.

Thus, even in the idealized situation in which a frequency splitting
could be translated into an intrinsic squared-mass splitting, the relevant
oscillatory component is not guaranteed to be observable: it may be
suppressed or absent depending on the interactions.

Let us also comment on tree level exchange diagrams.  In this case,
frequency-difference information, and hence potentially mass-splitting
information, can appear through a beat pattern between two mass-eigenstate
oscillations~\cite{Pinol:2021aun}.   Schematically, the
signal takes the form
\begin{equation}
  S_{\rm tree}(L)
 \sim
  A_1\cos\!\left(\mu_1L+\delta_1\right)
  +
  A_2\cos\!\left(\mu_2L+\delta_2\right).
  \label{eq:tree-beating}
\end{equation}
For equal amplitudes, \(A_1=A_2=A\), this can be rewritten as
\begin{align}
  S_{\rm tree}(L)
  \sim
  2A&
  \cos\!\left[
    \frac{\mu_1+\mu_2}{2}L
    +
    \frac{\delta_1+\delta_2}{2}
  \right]
  \notag\\
  &\times
  \cos\!\left[
    \frac{\mu_1-\mu_2}{2}L
    +
    \frac{\delta_1-\delta_2}{2}
  \right].
  \label{eq:tree-beat-product}
\end{align}
For \(\mu_1\simeq\mu_2\), the rapidly oscillating first factor is modulated
by a slowly varying envelope controlled by the mass difference.  More
generally, a clear beat pattern requires the two mass-eigenstate
contributions to have comparable amplitudes, $|A_1|\sim |A_2|$. 
If instead one contribution dominates, the signal is
approximately single frequency and the contrast of the beat modulation is
strongly reduced.  Even for comparable amplitudes, resolving the modulation
requires the accessible momentum range to cover an appreciable fraction of
the beat period.

The same conclusion applies more generally whenever two mass-eigenstate
oscillations contribute to the same squeezed-limit signal, including in
single-, double-, or triple-exchange topologies (see, e.g., Ref.~\cite{Qin:2023ejc} for exact analytic formulae
for single exchange and Ref.~\cite{Aoki:2024uyi} for double exchange).  The visibility of the
mass-difference modulation depends on their relative amplitudes, which are
controlled by the mixing parameters and interaction coefficients.
Consequently, the observability of such a modulation is model dependent
also at tree level.

These limitations motivate a complementary observable that uses the slow
time dependence of the Hubble parameter itself.  Instead of relying only on
mass splittings at a fixed Hubble scale, we now ask how a collider frequency
responds to the evolution of \(H\).

%----------------------------------------
\section{Hubble susceptibility}
\label{sec:hubble-susceptibility}
We now discuss how the slow evolution of the Hubble parameter can be used
to disentangle the intrinsic and Hubble-induced contributions to the
inflationary effective mass. 

For this purpose, we use \(N\equiv\ln a\), \(\epsilon\equiv-\dot H/H^2=-d\ln H/dN\),
and a prime for \(d/dN\). A dot denotes a cosmic time derivative.
We assume slow roll with \(\epsilon>0\), an adiabatically evolving scalar with standard dispersion,
\(\mu^2>0\) away from zero, and weak mixing.  Throughout this section
\(k_S\) is held fixed while \(k_L=k_S e^{-L}\) is varied, and
\(\partial_L\equiv(\partial/\partial L)_{k_S}\).

Within the minimal parametrization introduced
in Eq.~\eqref{m_eff0}, we regard the effective mass as a function of the
number of e-folds \(N\),
\begin{equation}
  m_{\rm eff}^2(N)
  =
  m_0^2+cH^2(N),\label{m_eff}
\end{equation}
where \(m_0^2\) and \(c\) are taken to be constant.  The corresponding
leading adiabatic cosmological collider frequency is
\begin{equation}
  \mu^2(N)
  =
  \frac{m_0^2}{H^2(N)}
  +c-\frac94 .
  \label{eq:mu-background}
\end{equation}

Although \(H\) is nearly constant during slow-roll inflation, its slow
evolution changes the relative importance of the intrinsic and
Hubble-induced mass contributions.  Differentiating
Eq.~\eqref{eq:mu-background} with respect to \(\ln H^2\) along the
inflationary background, we obtain
\begin{equation}
  \frac{m_0^2}{H^2}
  =
  -2\mu\frac{d\mu}{d\ln H^2},
  \qquad
  c
  =
  \mu^2+\frac94
  +2\mu\frac{d\mu}{d\ln H^2}.
  \label{diff}
\end{equation}
We refer to the response \(d\mu/d\ln H^2\) as the
\emph{Hubble susceptibility} of the collider mass.  Within the mass ansatz
above, the local frequency and its response to the Hubble scale therefore
allow the intrinsic and Hubble-induced contributions to be separated.

This response is encoded in the phase of the squeezed limit collider
oscillation.  For a slowly varying mass, the propagation
phase accumulated between the reference times of the long
and short modes can be written at leading order as
\begin{equation}
  \Theta(k_L,k_S)
  \simeq
  \int_{N_L}^{N_S}\mu(N)\,dN ,
  \label{eq:leading-phase}
\end{equation}
where \(N_L\) and \(N_S\) are defined by
\(k_L=a(N_L)H(N_L)\) and \(k_S=a(N_S)H(N_S)\), respectively.
For any background quantity \(f\), we write \(f_L=f(N_L)\),
\(f_S=f(N_S)\), and \(f'_{L,S}=(df/dN)|_{N_{L,S}}\).
These horizon-crossing times fix the phase convention. The finite
interaction regions are included in the matching phase below.
We use the leading slow-roll relation between scale and time
\begin{equation}
  N_S-N_L
  \simeq
  \ln\frac{k_S}{k_L}=L.
  \label{eq:leading-map}
\end{equation}

Expanding \(\mu(N)\) around \(N_S\),
\begin{equation}
  \mu(N)
  =
  \mu_S+\mu'_S(N-N_S)+\cdots ,
  \label{exp_mu}
\end{equation}
where $\mu'_S
  \equiv
  \left.\frac{d\mu}{dN}\right|_{N=N_S}$, we obtain
\begin{equation}
  \Theta(k_L,k_S)
  \simeq
  \mu_S L
  -\frac12\mu'_S L^2+\cdots.
  \label{eq:chirped-phase}
\end{equation}
At this order, \(\mu_S\) and \(\mu'_S\) are the linear and quadratic
coefficients of the phase expansion at fixed \(k_S\), over a range where
the omitted Taylor terms are controlled.  At a finite \(L\), the local
slope and curvature instead give \(\partial_L\Theta\simeq\mu_L\) and
\(\partial_L^2\Theta\simeq-\mu'_L\), so the same reconstruction below
applies locally with \(S\) replaced by \(L\).

Using \(d\ln H^2/dN=-2\epsilon\), we find
\begin{equation}
  \frac{d \mu}{d\ln H^2}
  =-\frac{\mu'}{2\epsilon}.
\end{equation}
Evaluating Eq.~\eqref{diff} at the short-mode reference time then gives
\begin{equation}
  \frac{m_0^2}{H_S^2}
  =\frac{\mu_S\mu'_S}{\epsilon_S},
  \qquad
  c=\mu_S^2+\frac94-\frac{\mu_S\mu'_S}{\epsilon_S}.\label{eq:m0_c_from_chirp}
\end{equation}
In this sense, the scale dependence of the collider oscillation measures
the Hubble susceptibility of the heavy mass.  Combining a fit to the local
frequency and its drift with independent information on \(H_S\) and
\(\epsilon_S\), one can in principle infer both the intrinsic mass
parameter \(m_0^2\) and the Hubble-induced correction \(c\) at leading order.  Such
information may be obtained, within a specified inflationary framework,
from tensor modes, their spectral observables, and the tensor-to-scalar ratio.

Although the chirp itself is a first order slow-roll effect, the leading
mass reconstruction does not require all other first order corrections to
the phase. The nonzero chirp retained here is
\(\mu'=\epsilon(m_0^2/H^2)/\mu\).
For a slow-roll counting parameter \(s\), with
\(\epsilon=O(s)\), \(\epsilon'=O(s^2)\), and
\(\mu,m_0^2/H^2=O(1)\), the frequency correction and the correction
to the mapping between scale and time shift the phase slope by \(O(s)\),
but its curvature only by \(O(s^2)\).
After division by \(\epsilon\), these give \(O(s)\) absolute corrections
to the reconstructed \(m_0^2/H^2\) and \(c\).
Equation~\eqref{eq:m0_c_from_chirp} therefore retains the leading
reconstruction. It does not require a phase prediction complete to
\(O(s)\).  The corrections are given in Appendix~\ref{app:phase-corrections}.

To visualize the effect, we consider the oscillatory part of a
representative tree level massive-exchange signal in the squeezed
limit.\footnote{
For definiteness, such a signal can be generated by the interactions
\begin{equation}
 S_{\rm int}=\int dt\,d^3\mathbf{x}\,a^3
 \left[\rho\,\sigma\dot\varphi
       +\frac{\sigma\dot\varphi^{\,2}}{2\Lambda}\right],
 \label{vertex}
\end{equation}
where the first term describes the linear mixing between the inflaton
fluctuation $\varphi$ and the massive field $\sigma$, while the second provides the cubic
vertex. The dot denotes a derivative with respect to the cosmic time $t$.  The fields are canonically normalized. \(\rho\) and
\(\Lambda\) are constant mass scales specifying the couplings.
We work at leading order in weak mixing, \(\rho/H\ll1\), and neglect
the \(O(\rho^2)\) shift of the oscillatory eigenfrequency.  If retained,
a finite constant \(\rho\) shifts the superhorizon squared mass to
\(m_{\rm eff}^2+\rho^2\), so that the Hubble-independent combination
inferred from the chirp becomes \(m_0^2+\rho^2\).}
Writing
\(x\equiv k_L/k_S=e^{-L}\ll1\), its squeezed limit nonanalytic part is approximated by
\begin{equation}
  S_{\rm osc}(k_S,x)
  \simeq
  \sqrt{x}\,{\cal A}_S(L)
  \cos[-\Theta(L)+\delta_S],
  \label{eq:chirped_template}
\end{equation}
where \({\cal A}_S(L)\equiv{\cal A}(k_S,L)\) is a slowly varying,
interaction-dependent amplitude, and \(\delta_S\) is an approximately
constant matching phase. Smooth matching-phase drift contributes only at
\(O(s^2)\) to the curvature under the same slow-roll counting, which must be included in a higher-precision fit. For \(\mu'_S=0\), choosing
\(\delta_S=-\mu_S\ln4-\pi/4\) reproduces the standard large mass template
\(S_{\rm osc}\propto \sqrt{x}\cos[\mu_S\ln(x/4)-\pi/4]\).

\begin{figure}[t!]
    \centering
    \includegraphics[width=8.5cm]{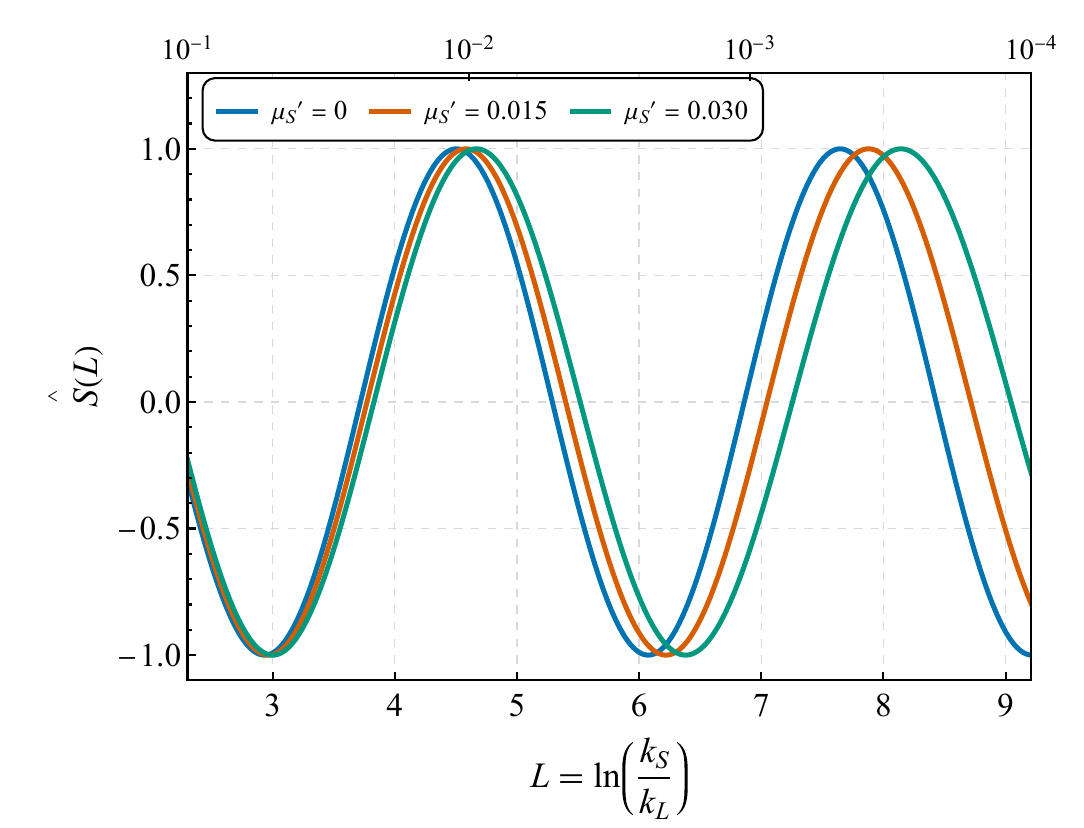}
   \caption{\it
Unit-normalized squeezed limit oscillation \(\widehat S(L)\) as a function of
\(L=\ln(k_S/k_L)\), for fixed \(k_S\), \(\mu_S=2\) and different drifts
\(\mu'_S\).  The upper ticks show the corresponding squeezed ratio
\(x=k_L/k_S=e^{-L}\).  A nonzero \(\mu'_S\) induces the quadratic phase
curvature in \(\Theta(L)=\mu_S L-\mu'_S L^2/2\), producing a chirp.
The curves are not a full bispectrum calculation, rather illustrate this truncated propagation phase with
the fixed matching phase \(\delta_S=-\mu_S\ln4-\pi/4\).
The drift values are illustrative and chosen to make the phase
modulation visible.
}
    \label{fig:Shape}
\end{figure}

In Fig.~\ref{fig:Shape}, we divide out the slowly varying amplitude and plot
the unit-normalized template
\begin{equation}
  \widehat S(L)
  \equiv
  \frac{S_{\rm osc}(k_S,e^{-L})}{e^{-L/2}{\cal A}_S(L)}
  \simeq
  \cos[-\Theta(L)+\delta_S],
  \label{eq:S_hat}
\end{equation}
with
\begin{equation}
  \Theta(L)
  \simeq
  \mu_S L-\frac12\mu'_S L^2 .
\end{equation}
We fix \(k_S\) and the expansion coefficient \(\mu_S\) and vary the drift \(\mu'_S\). Once \(\epsilon_S\) is independently determined, the
measured drift can be translated into \(m_0^2/H_S^2\) and \(c\) through
Eq.~\eqref{eq:m0_c_from_chirp}.

%----------------------------------------
\section{Conclusion}

Cosmological collider signals probe the effective masses of particles during
inflation, rather than directly measuring their Hubble-independent mass
parameters.  In this note, we have introduced the \emph{Hubble
susceptibility}, namely the response of a collider frequency to the slow
evolution of the Hubble scale.  This response is encoded in the scale
dependence of the squeezed limit oscillation: the local frequency determines
the effective mass, while the quadratic curvature of the accumulated phase
measures its drift.  For an effective mass consisting of a
Hubble-independent contribution and a term proportional to \(H^2\), these two
observables, together with background information on \(H\) and the slow-roll
parameter, allow the two mass contributions to be disentangled  at leading slow-roll and adiabatic order on a fixed-\(k_S\) slice.

We have also revisited the possibility of extracting Hubble-independent mass
information from mass splittings.  Although a difference-frequency signal can
remove a universal Hubble-induced contribution in an idealized situation, its
observability is itself interaction dependent.  
For the scalar bubble loop, the difference-frequency component is
suppressed near degeneracy for the representative time-derivative
interaction, whereas the nonanalytic oscillatory signal vanishes for
the non-derivative interaction in the exact de Sitter setting. At tree level,
mass-difference information can instead appear as a beat pattern, whose
visibility requires two mass eigenstate oscillations to contribute with
comparable amplitudes.  Thus, neither the existence of a nearly degenerate
spectrum nor a universal mass correction by itself guarantees an observable
mass-difference signal.

The Hubble susceptibility provides a complementary approach that does not
require multiple particles.  Its interpretation in terms of two separate
mass contributions relies on the assumed form of the inflationary effective
mass though additional time-dependent contributions generally make the
decomposition underdetermined.  Higher order scale dependence of the
collider phase can nevertheless provide consistency tests of this minimal
mass ansatz.  The main observational challenge is therefore to resolve the
small departure from exact log-periodicity in a primordial
cosmological collider signal.

%----------------------------------------
\vskip 4pt
\noindent\textbf{Acknowledgements.---}
We would like to thank Zhehan Qin and Yuhang Zhu for useful discussions. S.A. is supported by the Japan Science and Technology Agency (JST) 
as part of Adopting Sustainable Partnerships for Innovative Research Ecosystem (ASPIRE), 
grant JPMJAP2318, and by JSPS KAKENHI Grant Number 26K07078.
M.Y. is supported by IBS under the project code, IBS-R018-D3, and by
JSPS Grant-in-Aid for Scientific Research Number JP23K20843.

The authors used ChatGPT (OpenAI, GPT-5.6 Sol and GPT-6 Astra) and Claude
(Anthropic, Fable 5.1) to assist in checking derivations,
identifying possible inconsistencies, and improving the presentation
of the manuscript. All AI-assisted outputs were independently checked
by the authors, who take full responsibility for the content.
%----------------------------------------
\appendix

\section{Scalar-Bubble Signal from Exact Bubble Seeds}
\label{app}

We briefly revisit the two-scalar bubble contribution to the bispectrum
using the exact bubble seed~\cite{Xianyu:2022jwk,Qin:2024gtr}.  Throughout this appendix, we measure all masses and momenta in units of
the Hubble scale and set \(H=1\). We define the four-point master seed by
\begin{align}
  {\cal J}_{\mu_1,\mu_2}^{p_1p_2}
  \equiv
  &\,  k_s^{5+p_1+p_2}
  \sum_{a,b=\pm}(-ab)
  \int_{-\infty}^{0} d\tau_1 d\tau_2\,
  (-\tau_1)^{p_1}(-\tau_2)^{p_2}
  \notag\\
  &\times
  e^{ia k_{12}\tau_1+ib k_{34}\tau_2}
  {\cal B}_{ab}^{\mu_1,\mu_2}
  (k_s;\tau_1,\tau_2),
  \label{eq:bubble-seed}
\end{align}
where
\begin{align}
\nonumber   &{\cal B}_{ab}^{\mu_1,\mu_2}
  (k_s;\tau_1,\tau_2)\\
  &
  =
  \int\frac{d^3q}{(2\pi)^3}\,
  D_{ab}^{(\mu_1)}(q;\tau_1,\tau_2)
  D_{ab}^{(\mu_2)}
  (|\bm q+\bm k_s|;\tau_1,\tau_2).
\end{align}
Here \(a,b=\pm\) denote the Schwinger--Keldysh (SK) branches, and
\(D_{ab}^{(\mu_i)}\) is the massive scalar propagator with de Sitter
mass parameter $ \mu_i=\sqrt{m_{i,\rm eff}^2/H^2-9/4}$. We follow the SK conventions of
Ref.~\cite{Chen:2017ryl}.  The indices \(p_1\) and \(p_2\) characterize
the conformal-time dependence of the first and second vertices,
respectively.

We denote the magnitudes of the external momenta by
\(k_i\equiv|\bm k_i|\), and define $ k_{12}\equiv k_1+k_2, k_{34}\equiv k_3+k_4,
  k_s\equiv|\bm k_1+\bm k_2|$.
It is convenient to introduce the
dimensionless ratios
\begin{equation}
  r_1\equiv\frac{k_s}{k_{12}},
  \qquad
  r_2\equiv\frac{k_s}{k_{34}}.
  \label{eq:seed-ratios}
\end{equation}

The bispectrum contribution in which \( k_1\) and \(k_2\) are
attached to the first vertex is obtained by taking the single-folded
limit \(k_4\to0\) of the four-point seed.  Momentum conservation then
implies $k_s=k_3, k_{34}\to k_3$, or equivalently $ r_1=k_3/k_{12},
  r_2\to1^-$. We denote this permutation by \(S^{12|3}\). 
Up to overall coupling constants and analytic factors associated
with the external mode functions, it is generated from
\begin{equation}
  S^{12|3}
  \propto
  \left.
  {\cal D}_{\rm ext}\,
  {\cal J}_{\mu_1,\mu_2}^{p_1p_2}(r_1,r_2)
  \right|_{
    r_1=k_3/k_{12},\,
    r_2\to1^-
  }.
  \label{eq:seed-to-bispectrum}
\end{equation}
The operator \({\cal D}_{\rm ext}\) is analytic in the external momenta
and hence does not generate new nonanalytic frequencies.  It can,
however, modify the coefficients of the
\(\mu_1+\mu_2\) and \(\mu_1-\mu_2\) components.

For concreteness, we consider the schematic non-derivative interactions between inflaton fluctuation $\varphi$ and heavy scalars $\sigma_{1,2}$,
\begin{equation}
  {\cal L}_{\rm ND}^{(3)}
  \supset
  a^3 \,\varphi\,\sigma_1\sigma_2,
  \qquad
  {\cal L}_{\rm ND}^{(4)}
  \supset
  a^3\,(\varphi)^2\sigma_1\sigma_2 ,
  \label{eq:ND-interactions}
\end{equation}
and the representative time-derivative interactions
\begin{equation}
  {\cal L}_{\rm D}^{(3)}
  \supset
  a^3 \,\dot{\varphi}\,\sigma_1\sigma_2,
  \qquad
  {\cal L}_{\rm D}^{(4)}
  \supset
  a^3 \,(\dot{\varphi})^2\sigma_1\sigma_2 ,
  \label{eq:D-interactions}
\end{equation}
where the dot denotes a derivative with respect to cosmic time and coupling constants are suppressed.

After converting to conformal time, the non-derivative vertices carry
the factors \(a^4\), giving
\((p_1,p_2)=(-4,-4)\).  For the time-derivative vertices, using
\(\dot{\varphi}=a^{-1}\partial_\tau\varphi\) and
\(\partial_\tau\varphi_k\propto\tau e^{-ik\tau}\), the corresponding powers are
\((p_1,p_2)=(0,-2)\).  The bispectrum contributions are therefore
generated by
\begin{align}
  S_{\rm ND}^{12|3}
  &\propto
  \left.
  {\cal D}_{\rm ND}\,
  {\cal J}_{\mu_1,\mu_2}^{-4,-4}(r_1,r_2)
  \right|_{r_2\to1^-},
  \notag\\
  S_{\rm D}^{12|3}
  &\propto
  \left.
  {\cal J}_{\mu_1,\mu_2}^{0,-2}(r_1,r_2)
  \right|_{r_2\to1^-},
  \label{eq:representative-seeds}
\end{align}
where
\begin{equation}
  {\cal D}_{\rm ND}
  =
  \left(1-k_1\partial_{k_{12}}\right)
  \left(1-k_2\partial_{k_{12}}\right)
  \left(1-k_3\partial_{k_{34}}\right).
  \label{eq:D-ND}
\end{equation}
For the time-derivative contribution, the powers of conformal time arising
from the differentiated external mode functions have already been absorbed
into \((p_1,p_2)=(0,-2)\).  After suppressing analytic momentum
prefactors, no additional external-leg operator is required.

We extract the leading nonanalytic terms in the squeezed and folded
limits.  The compact expression below assumes infrared-safe time integrals. The
integer weights \(p_1,p_2\geq-2\) considered in
Ref.~\cite{Qin:2024gtr} cover the derivative choice \((0,-2)\), but
not the non-derivative choice \((-4,-4)\).
Within this domain, combining the local-signal and nonlocal-signal
branches before taking the folded limit, as in the scalar tree level
building blocks of Ref.~\cite{Aoki:2026vbc}, gives the following leading
nonanalytic expression, exact in the mass parameters. Continuation outside the domain requires a regulator.

We introduce
\begin{equation}
  \lambda_{s_1s_2}
  \equiv
  i\left(s_1\mu_1+s_2\mu_2\right),
  \qquad
  s_1,s_2=\pm1 ,
  \label{eq:lambda-sector}
\end{equation}
and
\begin{equation}
  {\cal A}_{s_1s_2}^{p_1p_2}
  \equiv
  \cos\frac{\pi(p_1-p_2)}{2}
  -
  \cos\left[
    \pi\lambda_{s_1s_2}
    +
    \frac{\pi(p_1+p_2)}{2}
  \right].
  \label{eq:A-sector}
\end{equation}
It is also useful to define the mass-dependent factor
\begin{align}
  {\cal G}_{s_1s_2}
  \equiv{}&
  \frac{\Gamma(-is_1\mu_1)\Gamma(-is_2\mu_2)}
       {\Gamma(3+\lambda_{s_1s_2})}
  \notag\\
  &\times\Gamma\left(\frac32+is_1\mu_1\right)
         \Gamma\left(\frac32+is_2\mu_2\right).
  \label{eq:G-sector}
\end{align}
The leading nonanalytic (signal) part of the folded seed is then
\begin{align}
\nonumber  \left.
  {\cal J}_{\mu_1,\mu_2}^{p_1p_2}
  \right|_{\rm sig}&\ \equiv   \left.
  {\cal J}_{\mu_1,\mu_2}^{p_1p_2}
  (r_1,r_2)
  \right|_{\substack{ r_1\to0,\,r_2\to1^-}}\\
  &\ 
  =
  \sum_{s_1,s_2=\pm1}
  {\cal J}_{s_1s_2}^{p_1p_2}(r_1),
  \label{eq:folded-signal-sum}
\end{align}
where
\begin{align}
  {\cal J}_{s_1s_2}^{p_1p_2}(r_1)
  =
  &\,
  \frac{
    2^{-6-p_2-\lambda_{s_1s_2}}
  }{\pi^3}\,
  r_1^{4+p_1+\lambda_{s_1s_2}}\,
  {\cal A}_{s_1s_2}^{p_1p_2}\,
  {\cal G}_{s_1s_2}
  \notag\\
\nonumber  &\times
\Gamma\left(4+p_1+\lambda_{s_1s_2}\right)
\Gamma\left(4+p_2+\lambda_{s_1s_2}\right)
     \\
  &\times \frac{\Gamma\left(1+p_2-\lambda_{s_1s_2}\right)}{\Gamma(3+p_2)}.
  \label{eq:folded-signal-sector}
\end{align}
It follows immediately that the four sectors scale as
\begin{equation}
  {\cal J}_{\pm\pm}^{p_1p_2}
  \propto
  r_1^{4+p_1\pm i(\mu_1+\mu_2)},
  \qquad
  {\cal J}_{\pm\mp}^{p_1p_2}
  \propto
  r_1^{4+p_1\pm i(\mu_1-\mu_2)},
  \label{eq:sum-difference-sectors}
\end{equation}
and therefore describe the sum- and difference-frequency components,
respectively.

For the representative time-derivative interaction,
\((p_1,p_2)=(0,-2)\), Eq.~\eqref{eq:folded-signal-sector}
is directly applicable.  The difference-frequency component is smaller than the sum-frequency
component and becomes parametrically suppressed near mass degeneracy. Defining
\begin{equation}
  \Sigma\equiv\mu_1+\mu_2,
  \qquad
  \Delta\equiv|\mu_1-\mu_2|,
\end{equation}
and denoting the coefficients of the sum- and difference-frequency
sectors by \({\cal C}_{+}^{\rm D}\) and \({\cal C}_{-}^{\rm D}\),
respectively, Eq.~\eqref{eq:folded-signal-sector} gives
\begin{equation}
  \frac{|{\cal C}_{-}^{\rm D}|}
       {|{\cal C}_{+}^{\rm D}|}
  =
  \frac{
    \sqrt{9+\Delta^2}\,
    \tanh\!\left(\frac{\pi\Delta}{2}\right)
  }{
    \sqrt{9+\Sigma^2}\,
    \tanh\!\left(\frac{\pi\Sigma}{2}\right)
  } .
  \label{eq:derivative-coefficient-ratio}
\end{equation}
Since
\(\sqrt{9+x^2}\tanh(\pi x/2)\) is monotonically increasing for
\(x>0\), and since \(\Delta<\Sigma\), the difference-frequency
coefficient is always smaller than the sum-frequency coefficient.
Moreover, in the nearly degenerate limit,
\begin{equation}
  \frac{|{\cal C}_{-}^{\rm D}|}
       {|{\cal C}_{+}^{\rm D}|}
  =
  \frac{
    3\pi
  }{
    2\sqrt{9+\Sigma^2}\,
    \tanh\!\left(\frac{\pi\Sigma}{2}\right)
  }
  \Delta
  +
  {\cal O}(\Delta^3).
  \label{eq:derivative-near-degenerate}
\end{equation}
Thus the magnitude of the difference-frequency coefficient vanishes
linearly as \(\Delta\to0\) at fixed \(\Sigma\), while the sum-frequency
coefficient remains finite. At fixed \(r_1\), the real sum of the two
conjugate difference-frequency sectors is further suppressed to
\(O(\Delta^2)\), since the leading term in their coefficients is purely
imaginary.
This analytically confirms, for the time-derivative contribution
considered here, the qualitative hierarchy found in the late-time
approximation~\cite{Aoki:2020zbj}.

For the non-derivative interaction with \((p_1,p_2)=(-4,-4)\), formal substitution in
Eq.~\eqref{eq:folded-signal-sector} gives \(1/\Gamma(3+p_2)=0\).
For fixed nonzero \(\lambda_{s_1s_2}\), this suggests a zero of the
displayed leading coefficient, but it does not establish the absence of the oscillatory bispectrum signal
outside the seed's infrared-safe domain.  In particular, regulating the vertex powers as
\(p_1=p_2=-4+\eta\), the difference-frequency sector
\(\lambda_{+-}=i(\mu_1-\mu_2)\) contains a factor proportional to
\(\eta/(\eta-\lambda_{+-})\), so the regulator and degenerate-mass
limits do not commute. The regulator, external-leg derivatives, folded limit, and
subleading nonanalytic terms must be treated consistently.
Thus the compact seed formula alone does not establish the scalar
non-derivative null.  We now show that the absence of the oscillatory
signal can nevertheless be established independently using the spectral
representation of the scalar bubble.

The scalar bubble admits a de Sitter K\"all\'en--Lehmann spectral
representation~\cite{Loparco:2023rug,Xianyu:2022jwk,Aoki:2026vbc},
\begin{equation}
  {\cal B}_{ab}^{\mu_1,\mu_2}
  (k;\tau_1,\tau_2)
  =
  \int d\nu\,
  \rho_{\mu_1,\mu_2}(\nu)\,
  D_{ab}^{(\nu)}(k;\tau_1,\tau_2),
  \label{eq:scalar-KL}
\end{equation}
where \(\rho_{\mu_1,\mu_2}(\nu)\) is the scalar-bubble spectral density
and \(D_{ab}^{(\nu)}\) is the SK propagator of a scalar with spectral
mass
\begin{equation}
  M_\nu^2=\frac94+\nu^2 .
\end{equation}
Possible local counterterms and background contributions do not affect
the nonanalytic oscillatory signal considered below. The bubble signal is therefore a spectral superposition of
tree level massive-exchange diagrams.

Upon inserting Eq.~\eqref{eq:scalar-KL} into the bubble diagram, the
non-derivative cubic and quartic vertices are mapped, for each spectral
component, onto a quadratic mixing and a cubic interaction involving the
spectral scalar, respectively. Keeping the observation time \(\tau_0<0\) finite, the corresponding mixed propagator
satisfies the identity~\cite{Aoki:2026vbc}
\begin{align}
\nonumber  & \sum_{b=\pm} ib
  \int_{-\infty}^{\tau_0} d\tau\,
  a^4(\tau)\,
  D_{ab}^{(\nu)}(k;\tau_1,\tau)\,
  G_{b+}(k;\tau,\tau_0)\\
  &
  =
  \frac{
    G_{a+}(k;\tau_1,\tau_0)
    -
    D_{a+}^{(\nu)}(k;\tau_1,\tau_0)
  }{M_\nu^2},
  \label{eq:mixed-propagator-scalar}
\end{align}
where \(G_{ab}\) denotes the massless inflaton propagator. This identity follows from the equations of motion and integration by
parts. It is first derived with both external times strictly earlier than
the SK turning point, and the observation time is then taken to that
turning point \(\tau_0\). With this prescription, the early-time boundary
term vanishes by the Bunch--Davies prescription, while the boundary terms
at the turning point cancel in the SK sum~\cite{Aoki:2026vbc}.

Thus each spectral tree exchange reduces to the difference between a
massless contact contribution and a contact diagram with one external
massive leg. The massless contact contribution is manifestly free of a
heavy field oscillatory signal.

For the second contribution, denoting the corresponding spectral scalar
by \(\sigma_\nu\), the non-derivative contact three-point function behaves
at late times as
\begin{equation}
  \left\langle
    \varphi_{\bm k_1}\varphi_{\bm k_2}\sigma_{\nu,\bm k_3}
  \right\rangle'_{\rm contact}
  \longrightarrow
  \frac{1}{4k_1^3k_2^3M_\nu^2},
  \qquad
  \tau_0\to0^- ,
\end{equation}
up to the overall coupling normalization.  Here the observation time
\(\tau_0\) is kept finite until the end of the calculation.   The surviving
contribution is therefore nonoscillatory. 

Hence every spectral component is free of the nonanalytic oscillatory
cosmological collider signal.  A subtlety is that, in general, the spectral
and time integrals need not be freely interchanged.  For the signal part,
however, this ordering issue does not affect the result, as shown in
Ref.~\cite{Aoki:2026vbc}, where the scalar case was explicitly checked.
We may therefore perform the spectral superposition at the level of the
cosmological-collider signal and obtain
\begin{equation}
  \left.
  S_{\rm ND}^{12|3}
  \right|_{\rm sig}
  =0 .
  \label{eq:scalar-ND-null}
\end{equation}
This statement concerns the nonanalytic oscillatory signal only. Local and
infrared-sensitive background contributions need not vanish.

The same null result can also be understood through the field
redefinition discussed in Ref.~\cite{Aoki:2026vbc}, which removes the
non-derivative quadratic mixing at this order up to local interactions.

%%%%%%%%%%%%%%%%%%%%%%%%%%%%%%
\section{Beyond the Minimal Mass Ansatz}
\label{app:general-mass}

In the main text, we assume that the effective mass takes the minimal form, Eq.~\eqref{m_eff}, where \(m_0^2\) and \(c\) are independent of time.  More generally, the mass may receive an additional contribution,
\begin{equation}
  m_{\rm eff}^2(N)
  =
  m_0^2+cH^2(N)+\Delta m^2(N).
  \label{eq:general-mass-app}
\end{equation}
The extra term may depend not only on \(H\), but also on other quantities
evolving along the inflationary background. In this case, Eq.~\eqref{eq:m0_c_from_chirp} is modified as 
\begin{align}
&\frac{m_0^2}{H^2}=\frac{\mu \mu^{\prime}}{\epsilon}-\frac{\Delta m^2}{H^2}-\frac{\left(\Delta m^2\right)^{\prime}}{2 \epsilon H^2},\\
&c=\mu^2+\frac{9}{4}-\frac{\mu \mu^{\prime}}{\epsilon}+\frac{\left(\Delta m^2\right)^{\prime}}{2 \epsilon H^2},
\end{align}
where a prime denotes a derivative with respect to \(N\). 

These expressions show that the phase parameters \(\mu\) and \(\mu'\) are
no longer sufficient to determine \(m_0^2\) and \(c\) separately.  In the
minimal ansatz~\eqref{m_eff}, the two phase observables, together with \(H\) and
\(\epsilon\), determine the two unknown mass parameters.  Once an arbitrary
contribution \(\Delta m^2(N)\) is allowed, however, the same observables
depend additionally on both the local value of \(\Delta m^2\) and its
time derivative.  The decomposition of the effective mass into
\(m_0^2\), \(cH^2\), and \(\Delta m^2(N)\) is therefore underdetermined
unless the additional contribution is independently known or restricted
to a specific parametrized form.

Additional phase information can nevertheless be used to test the minimal
mass ansatz.  Extending the phase expansion by one order in Eq.~\eqref{exp_mu},
\begin{equation}
  \Theta(L)
  \mathrel{\simeq}
  \mu_S L
  -\frac12\mu'_S L^2
  +\frac16\mu''_S L^3
  +\cdots ,
\end{equation}
the cubic phase coefficient determines \(\mu''_S\) within
the same leading fixed-\(k_S\) approximation. Locally at finite
\(L\), \(\partial_L^3\Theta\simeq\mu''_L\).  For
the minimal ansatz~\eqref{m_eff}, the frequency and its derivatives satisfy
\begin{equation}
  \mu'^2+\mu\mu''
  -
  \mu\mu'
  \left(
    2\epsilon+\frac{\epsilon'}{\epsilon}
  \right)
  =0 ,
  \label{eq:minimal-mass-consistency}
\end{equation}
which is the consistency relation that the minimal ansatz satisfies. Thus, provided that \(\epsilon\) and \(\epsilon'\) are independently known,
the cubic phase curvature supplies a consistency test of the minimal
two-component mass ansatz.

In the presence of the additional contribution, defining
\(F\equiv\Delta m^2/H^2\), the same combination becomes
\begin{equation}
  \mu'^2+\mu\mu''
  -
  \mu\mu'
  \left(
    2\epsilon+\frac{\epsilon'}{\epsilon}
  \right)
  =
  \frac12
  \left[
    F''
    -
    \left(
      2\epsilon+\frac{\epsilon'}{\epsilon}
    \right)F'
  \right].
  \label{eq:nonminimal-mass-consistency}
\end{equation}
A violation of Eq.~\eqref{eq:minimal-mass-consistency} would therefore
indicate time dependence beyond the minimal form \(m_0^2+cH^2\) once phase-extraction corrections at the required order have
been controlled.  A contribution constant in time or proportional to
\(H^2\) can instead be absorbed into \(m_0^2\) or \(c\).

\section{Propagation Phase and Slow-Roll Corrections}
\label{app:phase-corrections}

This appendix derives the leading phase relation used in the main text
and makes explicit the corrections omitted from its reconstruction.
For a canonically normalized scalar, negligible gradients and weak mixing
give the superhorizon equation
\begin{equation}
 \sigma''+(3-\epsilon)\sigma'
       +\frac{m_{\rm eff}^2}{H^2}\sigma=0.
 \label{eq:superhorizon-mode}
\end{equation}
Set \(\sigma=e^{-I}u\), where
\(I(N)=\tfrac12\int^N(3-\epsilon)\,d\widetilde N\).
Using \(I'=(3-\epsilon)/2\) and \(I''=-\epsilon'/2\), one obtains
\begin{align}
 u''+\omega^2u=0,\label{eq:canonical-mode}   
\end{align}
where
\begin{align}
\nonumber \omega^2&\equiv\frac{m_{\rm eff}^2}{H^2}
             -\frac{(3-\epsilon)^2}{4}+\frac{\epsilon'}2\\
&=\mu^2+\frac32\epsilon-\frac14\epsilon^2+\frac12\epsilon'.
 \label{eq:quasi-dS-frequency}
\end{align}
For \(\omega^2>0\), \(|\omega'|/\omega^2\ll1\), and
\(|\omega''|/\omega^3\ll1\), the WKB branches are
\begin{equation}
 u_\pm(N)\simeq\frac{1}{\sqrt{2\omega(N)}}
       \exp\!\left[\pm i\int^N\omega(\widetilde N)\,d\widetilde N\right].
 \label{eq:wkb-mode}
\end{equation}
Their propagation between the long- and short-mode reference times thus
supplies the phase
\begin{equation}
 \Theta(k_L,k_S)\equiv\int_{N_L}^{N_S}\omega(N)\,dN
 \label{eq:wkb-phase}
\end{equation}
at leading WKB order.  Taking \(\omega\simeq\mu\) gives
Eq.~\eqref{eq:leading-phase}.  The horizon-crossing endpoints are a
matching convention: gradients need not be negligible at the endpoints
themselves.  The corresponding matching phase must be included in the
observable combination \(-\Theta+\delta_{\rm match}\). The main-text
illustration approximates \(\delta_{\rm match}\) by \(\delta_S\).

More generally, let the positive monotonic functions \(\kappa_L(N)\)
and \(\kappa_S(N)\) specify separable reference conditions
\(k_L=\kappa_L(N_L)\), \(k_S=\kappa_S(N_S)\), and define
\(g(N)\equiv d\ln\kappa_L/dN\ne0\).
The notation of Sec.~\ref{sec:hubble-susceptibility} gives
\(\omega_L=\omega(N_L)\), \(g_L=g(N_L)\),
\(\omega'_L=(d\omega/dN)|_{N_L}\), and
\(g'_L=(dg/dN)|_{N_L}\).
At fixed \(k_S\), differentiation of
\(\ln\kappa_L(N_L)=\ln k_S-L\) yields
\begin{equation}
 \partial_LN_L=-\frac1{g_L},\qquad\partial_LN_S=0,
 \qquad\partial_L f_L=-\frac{f'_L}{g_L}.
 \label{eq:endpoint-chain-rule}
\end{equation}
The phase derivatives consequently are
\begin{equation}
 \partial_L\Theta=\frac{\omega_L}{g_L},\qquad
 \partial_L^2\Theta=-\frac{\omega'_Lg_L-\omega_Lg'_L}{g_L^3}.
 \label{eq:general-phase-derivatives}
\end{equation}
These are differentiation identities for the specified WKB phase.
The horizon-crossing prescription \(\kappa_{L,S}=aH\) gives
\begin{equation}
 g=1-\epsilon,\qquad
 L=\int_{N_L}^{N_S}(1-\epsilon)\,dN.
 \label{eq:horizon-map}
\end{equation}
The factor \(1-\epsilon\) follows from \(d\ln(aH)/dN\) and is
independent of the mass coefficient \(c\) in Eq.~\eqref{m_eff}.
Other interaction-time prescriptions require their own \(g\).

For ordinary slow roll, let \(\epsilon=O(s)\),
\(\epsilon'=O(s^2)\), with each additional background derivative
adding one power of \(s\), and \(\mu,m_0^2/H^2=O(1)\).
Equations~\eqref{eq:quasi-dS-frequency}--\eqref{eq:horizon-map} then give
\begin{align}
 \partial_L\Theta
 &=\mu_L+\epsilon_L\left(\mu_L+\frac{3}{4\mu_L}\right)+O(s^2),
 \label{eq:slope-order}\\
 \partial_L^2\Theta
 &=-\mu'_L+O(s^2),\qquad
 \mu'_L=\frac{\epsilon_L}{\mu_L}\frac{m_0^2}{H_L^2}.
 \label{eq:curvature-order}
\end{align}
The \(O(s)\) correction to the slope therefore does not change the
leading \(O(s)\) chirp.  In particular, at finite squeezing,
\begin{align}
 &-\frac{(\partial_L\Theta)(\partial_L^2\Theta)}{\epsilon_L}
 =\frac{m_0^2}{H_L^2}+O(s),\label{eq:finite-mass-reconstruction}\\
 &(\partial_L\Theta)^2+\frac94
 +\frac{(\partial_L\Theta)(\partial_L^2\Theta)}{\epsilon_L}
 =c+O(s).
 \label{eq:finite-c-reconstruction}
\end{align}
This is the accuracy claimed for the main-text reconstruction.  It
retains \(\mu'\) and does not set \(\epsilon\) to zero in the mass drift.

For illustration of the subleading terms, suppose \(\epsilon\) is
constant and nonzero and use the horizon-crossing prescription.
Defining \(p\equiv\partial_L\Theta\) and
\(q\equiv\partial_L^2\Theta\), Eq.~\eqref{eq:general-phase-derivatives}
gives \(\omega_L=(1-\epsilon)p\) and
\(\omega'_L=-(1-\epsilon)^2q\).  Since
\(\omega_L\omega'_L=\epsilon m_0^2/H_L^2\), the inversion becomes
\begin{align}
 &\frac{m_0^2}{H_L^2}
 =-\frac{(1-\epsilon)^3}{\epsilon}\,p q,
 \label{eq:constant-epsilon-mass}\\
 &c=(1-\epsilon)^2p^2+\frac{(3-\epsilon)^2}{4}
       +\frac{(1-\epsilon)^3}{\epsilon}\,p q.
 \label{eq:constant-epsilon-c}
\end{align}
These equalities hold within the specified leading WKB phase.
They do not, by themselves, provide improved estimators for the full
signal phase. A consistent correction at this order also requires the
matching phase. For varying \(\epsilon\), the \(g'\) and \(\epsilon'\) terms above and
the derivative of Eq.~\eqref{eq:quasi-dS-frequency} must be retained.
A precision fit also includes the matching phase: if it depends smoothly
on slow background parameters, its slope is \(O(s)\) and curvature
\(O(s^2)\), so it likewise first affects the mass inversion at \(O(s)\).
Rapidly varying interactions or backgrounds fall outside this counting.
Finally, the constant-\(c\) ansatz itself can receive slow-roll
corrections. For example, a constant non-minimal curvature coupling gives
\(\xi R=6\xi(2-\epsilon)H^2\) in spatially flat FLRW spacetime.

%----------------------------------------
\bibliographystyle{apsrev4-1}
\bibliography{Refs}

\end{document}